\documentclass{elsart}

\usepackage{amsmath}
\begin{document}
\begin{frontmatter}
\title{Quantum teleprotation with sonic black holes }
\author[shao,nao,gs]{Xian-Hui Ge}
\author[shao,nao,itp,email]{You-Gen Shen}
\address[shao]{Shanghai Astronomical Observatory,Chinese Academy of Sciences,
Shanghai 200030, China (mail address)}
\address[nao]{National Astronomical Observatories, Chinese Academy of Sciences,
Beijing 100012, China}
\address[gs]{Graduate School of Chinese
Academy of Sciences, Beijing 100039, China}
\address[itp]{Institute of Theoretical Physics, Chinese Academy of Sciences,
Beijing 100080, China}
\thanks[email]{e-mail:gexh@shao.ac.cn\\ ygshen@shao.ac.cn}

\begin{abstract}
 We show a new property of sonic black holes. After deriving
  the metric of a sonic black hole from the
Schr$\ddot{o}$dinger equation and quantizing the perturbation
fields near the sonic event horizon, we show particles of Hawking
radiation can act as a source of entanglement: two-mode squeezed
entanglement is produced near the event horizon, which can be used
in quantum teleportation. The fidelity of the teleportation is
closely related to the temperature of the sonic black holes, but
high fidelity seems difficult to reach in our case.
\end{abstract}
\end{frontmatter}
\newpage
\hspace*{7.5mm} Recent investigation in string theory indicates
that black hole may not actually destroy the information about how
they were formed, but instead process it and emit the processed
information as part of the Hawking radiation as they
evaporate[1-5]. Lloyd suggested that [6], if this is correct, then
black holes can act as quantum computers: when a black hole is
formed, information is encoded in the initial
 conditions to be processed by the planckian
dynamics at the hole's horizon, and answers can be extracted to
the computation by examining the correlation in the Hawking
radiation emitted when the hole evaporate. Lloyd's proposal is
supported by Horowitz and Maldecena's (HM) theory of "the black
hole final state"[7]. HM have proposed a simple model of black
hole evaporation by imposing a final boundary condition near the
black hole singularity. The annihilation of infalling particles
and the collapsed matter can act as a measurement, transferring
the information contained in the matter to the outgoing Hawking
radiation.  Although interesting, HM's conjecture seems fall
outside the usual formulation of quantum mechanics and was
criticized by several authors[8-11]. Moreover, whether Hawking
radiation can transfer information also remains as an
enigma to most people. \\
\hspace*{7.5mm} In 1981, Unruh developed a way of mapping certain
aspects of black holes in supersonic flows and pointed out that
propagation of sound in a fluid or gas turning supersonic[12], is
similar to the the propagation of a scalar field close to a black
hole, and thus experimental investigation of the Hawking radiation
is possible.  From then on, several candidates have been
considered for the experimental test of the sonic analog of black
holes, which include superfluid helium II[13], atomic
Bose-Einstein condensates[14], one-dimensional (1D) transonic
flows[15], dielectric[16], wave gravity[17], and 1D
Fermi-degenerate noninteracting gas[18]. Similar to classical
black holes, sonic black holes also have the structure of
ergosphere, trapped regions, apparent horizon, and event
horizon[14]. The difference is that sonic black hole has no
singularity.  In this paper, we will demonstrate quantum
teleportation can be realized by using the squeezed vacuum states
produced from a sonic black hole's horizon. The fidelity of the
teleportation is closely
related to the temperature of a sonic black hole. \\
 \hspace*{7.5mm}Since the discussions on the Hawking radiation in sonic black
 holes have been extended to quantum fluids, we will first derive
 the metric of a sonic black hole and start from the
 Schr$\ddot{o}$dinger equation, which can result in the
 (barotropic, inviscid, and irrotational) hydrodynamics equations[19]. Then we will quantize
 the fluctuation field and see how quantum teleportation is possible with a
 sonic black hole.
  The non-relativistic quantum
 mechanics can be described by the Schr$\ddot{o}$dinger equation
  $[-\frac{\hbar ^2}{2m}\nabla ^2 +V(\vec{x})]\psi
   (\vec{x},t)=i\hbar\frac{\partial}{\partial t}\psi (\vec{x},t)$,
   where $V(\vec{x})$ is an effective potential, and $\psi (\vec{x},t)$ might be considered as a
   macrostate of condensate matter. The Lagrangian
   corresponding to the Schr$\ddot{o}$dinger equation can be
   defined as $L= \frac{i\hbar}{2}(\psi ^{*}\frac{\partial \psi}
   {\partial t}-\psi \frac{\partial \psi^{*}}{\partial t})-
   \frac{\hbar^2}{2m}\nabla \psi ^{*}\cdot \nabla
   \psi-V(\vec{x})\psi^{*}\psi$, which can be further rewritten as
   \begin{equation}
   L=-\psi^{*}\psi\{\frac{\hbar}{2i}\frac{\partial}{\partial t}ln\psi/\psi ^{*}+
   \frac{\hbar^2}{2m}\nabla(ln\psi ^{*})\nabla(ln\psi ^{*})+V(\vec{x})\}.\end{equation}
   Comparing Eq.(1) with the Jacobi-Hamilton equation, i.e.
   \begin{equation}
   \frac{\partial S}{\partial t}+\frac{(\nabla S)^2}{2m}+V(\vec{x})=0,\end{equation}
   where S is the action of the whole system, one obtains
   $S=S^{*}=\frac{\hbar}{i}ln\psi=-\frac{\hbar}{i}ln\psi^{*}$.
   Assumed $S=S_{r}+i S_{i}$, $\psi$ can be rewritten as \begin{equation}
   \psi (\vec{x},t)=e^{\frac{i(S_{r}+i S_{i})}
   {\hbar}}=\rho^{1/2} (\vec{x},t)e^{\frac{i
   S_{r}(\vec{x},t)}{\hbar}},\end{equation}
   where $\rho=\psi\psi^{*}$ is  the probability density.
   Substituting Eq.(3) into the Schr$\ddot{o}$dinger equation, we
   have
   \begin{eqnarray}
   \frac{\partial \rho}{\partial t}+\nabla \cdot
   \vec{j}(\vec{x},t)=0,\\
   \frac{\partial S_{r}}{\partial t}+\frac{(\nabla S_{r})^2}{2m}+V(\vec{x})
   -\frac{\hbar^2}{2m}\frac{\nabla ^2 \rho^{1/2}}{\rho^{1/2}}=0,\end{eqnarray}
   where $\vec{j}(\vec{x},t)=\frac{\rho}{m}\nabla S_{r}$ and the last term in Eq.(5)
    corresponds to the quantum effect without
    classical correspondence.
     We may drop out this
   term in that $\hbar^2$ is small and $\rho\leq 1$.
   If we set $\vec{v}=\nabla S_{r}/m$, then Eq.(5) can be
   rewritten as \begin{equation} \frac{\partial \vec{v}}{\partial t}
   +\nabla(\vec{v}^2/2)=-\nabla V(\vec{x})/m,\end{equation}
   which is the exact equation of irrotational fluid.
   Defining $\Phi=S_{r}/m$ and
   $V(\vec{x})=\phi(\vec{x})+\int^{p}_{0}
   d p'/\rho(p')
   $, we have \begin{equation}\frac{\partial \Phi}{\partial t}+\vec{v}^2/2=
   \frac{\phi(\vec{x})}{m}-\frac{1}{m}\int^{p}_{0}
   d p'/\rho(p'),\end{equation}
   where $p'$ denotes the pressure and $\phi(\vec{x})$ denotes the external potential. By further defining $\xi=ln \rho$ and $g(\xi)=\int^{p}_{0}
   d p'/\rho(p')$, we have the new forms of Eq.(4) and Eq.(7)
   \begin{eqnarray}\partial \xi/\partial
   t+\vec{v}\cdot\nabla\xi+\nabla \cdot \vec{v}=0,\\
   \frac{\partial \Phi}{\partial
   t}+\vec{v}^2/2-\frac{\phi(\vec{x})}{m}+g(\xi)/m=0.\end{eqnarray}
    Linearizing
   Eq.(8) and Eq.(9) around the assumed background ($\xi_{0}$,
   $\Phi_{0}$), with $\xi=\xi_{0}+\tilde{\xi}$ and
   $\Phi=\Phi_{0}+\tilde{\Phi}$, we obtain \begin{equation}
   \rho_{0}^{-1}[\partial \rho_{0}\tilde{\xi}/\partial
   t+\nabla\cdot (\rho_{0}\vec{v}\tilde{\xi}+\rho_{0}\nabla \tilde{\Phi}]=0,
   \partial \tilde{\Phi}/\partial t+\vec{v}\cdot\nabla\tilde{\Phi}
   +\frac{1}{m}g'(\xi_{0})\tilde{\xi}=0,\end{equation}
   which result in an equation for $\tilde{\Phi}$,

   \begin{equation}\rho_{0}^{-1}\left\{\frac{\partial}{\partial t}
   \left(\frac{m \rho_{0}}{g'(\xi_{0})}\frac{\partial \tilde{\Phi}}{\partial t}
   +\frac{m\rho_{0}\vec{v}_{0}}{g'(\xi_{0}) }\cdot \nabla\tilde{\Phi}\right)
   +\nabla\cdot\left(\frac{m\rho_{0}\vec{v}}{g'(\xi_{0})}\frac{\partial \tilde{\Phi}}{\partial t}
   -\rho_{0}\nabla \tilde{\Phi} +\vec{v}\frac{m\rho_{0}\vec{v}}{g'(\xi_{0})}
   \vec{v}\cdot\nabla \tilde{\Phi}\right)\right\}=0,\end{equation}
   which describes the propagation of the a massless scalar field
   in terms of metric.
   The local speed of sound is defined by $c^2\equiv
   g'(ln\rho_{0})/m$, which is assumed to be a constant. The
   metric can be written as $ds^2=\rho_{0}/c[(c^2-v_{0}^2)dt^2+2\vec{v}_{0}\cdot d\vec{x}dt-d\vec{x}\cdot
   d\vec{x}]$. Assuming that the background flow is a spherically
   symmetric,stationary, and convergent flow,we can define a new
   time coordinate by $d\tau=dt+\frac{\vec{v}_{0}\cdot
   d\vec{r}}{c^2-v^2}$. Substituting this back into the line
   element gives
   \begin{equation}
   ds^2=\frac{\rho_{0}}{c}\left[(c^2-v_{0}^2)d\tau^2-\frac{c^2}{c^2-v_{0}^2}dr^2
   -r^2(d\theta^2+sin\theta^2d\varphi^2)\right]\end{equation}
   As suggested by Unruh that at $r=R$, we assume the background
   fluid smoothly exceeding the velocity of sound, $v_{0}=-c+\alpha
   (r-R)+\textit{O}((r-R)^2)$. Then the  metric can be reexpressed
   as
   \begin{equation}ds^2=\frac{\rho_{0}}{c}\left(2c\alpha(r-R)d\tau^2-
   \frac{cdr^2}{2\alpha(r-R)}-r^2(d\theta^2+sin\theta^2d\varphi^2)\right),\end{equation}
   which has the same form of the Schwarzschild metric. Dropping the angular part of the metric will not
   change our results in the following discussion.\\

When we quantize the field $\tilde{\Phi}$, we will find the
behavior of the normal modes near the sonic horizon implies that
this sonic black hole will emit sound waves with a thermal
spectrum and particles will be produced near the horizon. The
particles created from the vacuum near the horizon are actually
the modes of squeezed vacuum states, which are approximate to the
EPR (Einstein-Podolsky-Rosen) states and can be used in quantum
teleportation. A distant observer from the horizon ($r\gg R$) can
be regarded as an observer in Minkowski space. Thus, a massless
scalar quantum field $\tilde{\Phi}$ in the D-dimensional Minkowski
space
     -time can be decomposed in Minkowski modes $\{U_{k}(x)\}$,
     which goes as
     \begin{equation}\tilde{\Phi}=\sum_{k}\left[a_{k}U_{k}(x)+{a}_{k}^{\dagger}U_{k}^{\ast}(x)\right],\end{equation}
     where ${a}_{k}$, ${a}_{k}^{\dagger}$ are annihilation and creation operators
     respectively,
     the boundary conditions $k_{n}^{i}=2\pi L_{i}^{-1}n_{i}$ ($i=1,...,D$) and $k=(k_{1},\vec{k})$.
     The Minkowski vacuum can be defined by
     \begin{equation}a_{k}\mid 0_{M}>=0, \forall  k.\end{equation}
By solving the Klein-Gordon equation in the coordinates of Eq.
(13), the field $\tilde{\Phi}$ can be expanded in normal modes:
\begin{equation}\tilde{\Phi}=\sum_{\sigma}\sum_{p}\left[b_{p}^{(\sigma)}u_{p}^{(\sigma)}(x)
+{b}_{p}^{(\sigma)\dag}u_{p}^{(\sigma)*}(x)\right],\end{equation}
 where the operators $b_{p}^{(\sigma)}$ and ${b}_{p}^{(\sigma)\dagger}$
 are assumed to satisfy the usual canonical communication
 relations, $p=(\Omega,\vec{k})$, and the symbol $\sigma=\pm$ refers to region I and II respectively, which is
 separated by the event horizon. By introducing the Unruh modes[20]
 \begin{equation}d_{p}^{(\sigma)}=\int_{-\infty}^{\infty} d\vec{k} p_{\Omega}^{(\sigma)}
 (\vec{k})a_{k_{1},\vec{k}},\end{equation}
  where $\{p_{\Omega}^{(\sigma)}(\vec{k})\}$
 is the complete set of orthogonal functions. The  modes $b_{p}^{(\sigma)}$ and ${b}^{(-\sigma)}_{\tilde{p}}$
 can be well expressed in terms of the Unruh modes by the Bogolubov
 transformations
 \begin{eqnarray}
 b^{(\sigma)}_{p}=[2sinh(2\omega \pi/\alpha)]^{-\frac{1}{2}}\left[e^{\omega \pi/\alpha}
 d^{(\sigma)}_{p}+e^{-\omega \pi/\alpha}{d}^{(-\sigma)\dagger}_{ \tilde{p}}\right],\\
 {b}^{(-\sigma)}_{\tilde{p}}=[2sinh(2\omega \pi/\alpha)]^{-\frac{1}{2}}\left[e^{-\omega \pi/\alpha}
 d^{(\sigma)\dagger}_{ p}+e^{\omega \pi/\alpha}{d}^{(-\sigma)}_{\tilde{ p}}\right],\end{eqnarray}
 where $\tilde{p}=(\Omega,-\vec{k})$. One then obtains[20]
\begin{equation}
\mid0>_{M}={Z} \prod_{\sigma,p}exp(tanhr
b_{p}^{(\sigma)\dagger}{b}_{\tilde{p}}^{(-\sigma)\dagger})\mid0_{I}>\otimes
|0_{II}>,\end{equation} where Z is a normalization constant
$Z=\prod_{p}cosh^{-2}r$, where $r=r(p)$,
 $tanhr=e^{-\omega \pi/ \alpha}$, and $coshr=(1-e^{-2\omega \pi/\alpha})^{-1/2}$.
  Assumed \begin{eqnarray}
T^{(+)}(r)=-\sum_{p}(b_{p}^{(+)\dagger}b_{p}^{(+)}lnsinh^{2}r-b_{p}^{(+)}
b_{p}^{(+)\dagger}lncosh^{2}r
),\nonumber\\
T^{(-)}(r)=-\sum_{p}(b_{\tilde{p}}^{(-)\dagger}b_{\tilde{p}}^{(-)}
lnsinh^{2}r-b_{\tilde{p}}^{(-)}b_{\tilde{p}}^{(-)\dagger}lncosh^{2}r
),\end{eqnarray} the following can be easily proved
\begin{eqnarray}
e^{-T^{(\sigma)}(r)/2}b_{p}^{(\sigma)\dagger}e^{-T^{(\sigma)}(r)/2}=
b_{(p)}^{(\sigma)
\dagger}tanhr,\nonumber\\
e^{-T^{(\sigma)}(r)/2}|0>_{I,II}=\prod_{p}cosh^{-1}r\mid0_{I}>\otimes
|0_{II}>.\end{eqnarray}  By using Eqs.(22), one can rewrite
Eq.(20) as
\begin{eqnarray}
|0>_{M}&=&\left(e^{-T^{(\sigma)}/2}e^{\sum_{p}b_{p}^{(\sigma)
\dagger}b_{\tilde{p}}^{(-\sigma)}}e^{T{(\sigma)}/2}\right)\left(e^{-T^{(\sigma)}/2}|0>_{I,II}\right)\nonumber\\
&=&e^{\sum_{p}\left[n_{p}lnsinhr-(1+n_{p})lncoshr\right]}
\sum_{n_{p}=0}^{\infty}|n_{p}>_{I}\otimes|n_{\tilde{p}}>_{II}\nonumber\\
&=&\sum_{n_{p}=0}^{\infty}\prod_{p}tanh^{n_{p}}rcosh^{-1}r|n_{p}>_{I}\otimes
 |n_{\tilde{p}}>_{II},
\end{eqnarray}where $N_{p}=b_{p}^{(\sigma)\dagger}b_{p}^{(\sigma)}$, $b_{p}^{(\sigma)}b_{p}^{(\sigma)\dagger}=1+N_{p}$.
 From Eq.(17),
we see that a given Minkowski mode of frequency $\omega_{\vec{k}}$
is spread over all positive frequencies $\Omega$, as a result of
the Fourier transform relationship between $a_{k_{1},\vec{k}}$ and
$d^{(\sigma)}_{p}$[21]. In fact, only one mode produced by the
sonic horizon is enough for our purpose. Thus, one can only
consider the mode p in region I and the mode $\tilde{p}$ in region
II. Therefore, the single mode component of the Minkowski vacuum
state, namely the two-mode vacuum squeezed state can be given by
\begin{equation}
\mid 0>_{M}=\frac{1}{cosh r}\sum^{\infty}_{n=0}
tanh^{n}r|n>_{I}\otimes|n>_{II}.
\end{equation}
 Equation (24) demonstrates that the vacuum appears as an entangled
 state of Hawking particles with nonlocal EPR type
 correlations and the vacuum can approach a perfectly entangled state
 of particle number asymptotically in the limit
 $tanhr\rightarrow 1$. The reduced density of each mode is a thermal-like
  state with mean particle number $\bar{n}=\frac{1}{e^{2\pi \omega/\alpha}-1}$,
  with the temperature $T=\frac{\hbar\alpha}{2\pi k}$, which is the Hawking
  temperature of the sonic black hole. As the above two-mode squeezed vacuum state is
  produced by the sonic horizon, we can set up the teleportation protocol: suppose
  one mode is open to local operations and measurements at the sender's location A
  by observer Alice, while the other mode is open to local operations and measurements
  in the receiver's location B, by observer Bob. Alice and Bob can communicate with each other
  via a classical channel. In the protocol, we consider the target mode $T$,
  in an unknown state $|\varphi>_{T}$.  The input state of the complete
system is
\begin{equation}cosh^{-1}r \sum_{k=1}^{\infty}tanh^{k}r|\varphi_{T}>\otimes|k_{I}>\otimes|k_{II}>\end{equation}
The teleportation protocol here require Alice make measurements on
T and I by using a unitary operator $U$, which can be defined as
$U=S_{1}\otimes S_{2}$. Suppose the combined system state after
Alice's measurement is $|\Theta>\otimes |X_{II}>$, where
$|X_{II}>$ is the state of Bob after the teleportation and
$|\Theta>=\gamma \sum_{j=1}^{\infty}|j_{T}
>\otimes |j_{A}
>$, $\gamma$ is the normalization constant and $|j_{T}$($j=1,2,...$) are the fixed orthonormal
bases for the Hilbert space of the target state $\varphi_{T}$.
Thus, we have
\begin{eqnarray}
|X_{II}>&=& \gamma cosh^{-1}r\sum_{j=1}^{\infty} <j_{T}|\otimes
<j_{A}|(S_{1}\otimes S_{2}) \otimes
\sum_{k=1}^{\infty}tanh^{k}r|\varphi_{T}>\otimes|k_{I}>\otimes|k_{II}>
\nonumber\\
 &=& \gamma cosh^{-1}\sum_{j=1}^{\infty}\sum_{k=1}^{\infty}tanh^{k}r <j_{T}|S_{1}|\varphi_{T}>
 <j_{A}|S_{2}|k_{I}>|k_{II}> \end{eqnarray}
 In terms of the basis components $X_{II j}=<j_{II}|X_{II}>$,
 $\varphi_{k}=<k_{T}|\varphi_{T}>$, $S_{1 jk}=<j_{T}|S_{1}|k_{T}>$ and
 $S_{2 jk}=<j_{I}|S_{2}|k_{I}>$, we have
 \begin{equation}X_{II j}=\gamma cosh^{-1}r \sum_{l=1}^{\infty}
 tanh^l r(S_{2}^{T}S_{1})_{jl}\varphi_{T l},\end{equation}
 where $S^{T}$ denotes matrix transpose of S. After Alice's
 measurements, she can inform Bob what the unitary transform
 $S_{2}^{T}S_{1}$ is via the classical channel.
 The original work on teleportation with squeezed vacuum state
 has been done by  Milburn and Braunstein (MB) in 1999 [22,23].
 They pointed out that the entanglement of Eq.(24) can be viewed
 in two ways: as an entanglement between quadrature phases  or as
 an entanglement between number and phase in the two modes[22].
 In case of teleportation using number and phase measurements,
 the modes obtained by Bob can be expressed as[22]
 \begin{equation}
 |X^{(\pm k)}>=\frac{(1-tanh^2r)^{1/2}}
 {\sqrt{P_{\pm}(k)}}\sum_{n=2k,0}^{\infty}tanh^{n\pm 2k}r c_{n}|n\pm 2k>_{II},\end{equation}
 where $P_{\pm}(k)=(1-tanh^{2}r)\sum_{n=0,2k}^{\infty}\lambda ^{2n}|c_{n\pm
 2k}|^2$.
 MB's results show that when the mean particle number in the
 entanglement resource is significantly greater than in the target
 state, the teleportation  has high fidelity. In
 particular, if the target state is a coherent state with
 amplitude $\varsigma$, the fidelity for zero particle-number
 difference measurement is[22]
 \begin{equation}
 F(0)=e^{-|\varsigma|^2 (1-tanhr)^2}.\end{equation}
 We see that if $tanhr\rightarrow 1$ and the photon-number is
 zero, then the teleportation is perfect, which infers that
 $\alpha\rightarrow\infty$. However, considered the velocity can not change greatly when liquid pass
 the horizon, $\alpha\rightarrow\infty$
 is not permitted. Therefore, perfect teleportation seems impossible
 by using the two-mode squeezed vacuum state from a sonic black
 hole horizon.\\

 In summary, we have investigate the possibility of using particles of Hawking
 radiation from sonic black hole horizon as a source of quantum
 teleportation. The fidelity of teleportation is shown rely on
 the the Hawking temperature in that $T\propto \alpha$.
  Considered the recent report on Hawking temperature of sonic
  black holes($\sim 200 nK$)[18], perfect teleportation seems impossible to reach
  by using sonic black holes.

 \textbf{Acknowledgements}\\ The work has been supported by the National
Natural Science Foundation of China under Grant No. 10273017.

\end{document}